\documentclass[12pt]{article}
\usepackage{a4wide}
\usepackage[centertags]{amsmath}
\begin{document}
\begin{flushright}
\begin{tabular}{l}
UWThPh-1996-33
\\
hep-ph/9605305
\\ \\
\end{tabular}
\end{flushright}
\def\thefootnote{\fnsymbol{footnote}}
\begin{center}
\huge
The problem of the spin of the proton
and elastic neutrino-proton scattering\footnote{
Talk presented by S.M. Bilenky
at the Adriatico Research Conference on
\textit{Trends in Collider Spin Physics},
ICTP, Trieste, December 1995.
}
\\
\vspace{0.5cm}
\Large
S.M. Bilenky
\\
\vspace{0.2cm}
\normalsize
Joint Institute for Nuclear Research, Dubna, Russia,\\
and\\
Institut f\"ur Theoretische Physik,
Universit\"at Wien,\\
Boltzmanngasse 5,
A-1090 Vienna, Austria\\
\vspace{0.2cm}
\large
and\\
\vspace{0.2cm}
\Large
W.M. Alberico,
C. Giunti,
C. Maieron
\\
\vspace{0.2cm}
\normalsize
INFN, Sezione di Torino,
and Dipartimento di Fisica Teorica, Universit\`a di Torino,
\\
Via P. Giuria 1, 10125 Torino, Italy
\end{center}

The experiments on the measurement
of the deep inelastic scattering
of longitudinally polarized leptons
on longitudinally polarized nucleons\cite{EMC,CERN,SLAC}
led to an essential progress in the investigation
of the structure of the nucleon.
The recent experiments
done at CERN\cite{CERN}
and at SLAC\cite{SLAC}
confirm the conclusion that have been reached
after the EMC data\cite{EMC}
became available:
the one-nucleon matrix element
of the strange axial current
is large and it's size is comparable
with the matrix elements of the $u$ and $d$
axial currents.
From the analysis of the latest data
it follows that\cite{Ellis}
\begin{equation}
\begin{array}{l} \displaystyle
g_{A}^{u}
=
0.83 \pm 0.03
\;,
\\ \displaystyle
g_{A}^{d}
=
- 0.43 \pm 0.03
\;,
\\ \displaystyle
g_{A}^{s}
=
- 0.10 \pm 0.03
\;.
\end{array}
\label{06}
\end{equation}
The constants $g_{A}^{q}$
are determined by
\begin{equation}
\left\langle
p
\left|
\bar{q} \gamma_{\alpha} \gamma_{5} q
\right|
p
\right\rangle
=
\bar{u}(p)
\,
\gamma_{\alpha} \, \gamma_{5}
\,
u(p)
\,
g_{A}^{q}
\;,
\qquad
q=u,d,s
\;,
\label{01}
\end{equation}
where
$ \left| p \right\rangle $
is the state vector of a nucleon
with momentum $p$.
The values (\ref{06})
of the constants
$g_{A}^{q}$
were determined under the assumption
of a Regge behaviour
of the polarized structure function
$g_1$ at small $x$.
The SU(3)-based relation
between the constants
$g_{A}^{q}$
and the constants $F$ and $D$
were also used.

Alternative approaches to the
``problem of the spin of the proton''
are of great interest.
The investigation of the NC-induced processes
\arraycolsep=0pt
\begin{eqnarray}
&&
\nu_\mu + N \to \nu_\mu + N
\label{091}
\\
&&
\bar\nu_\mu + N \to \bar\nu_\mu + N
\label{092}
\end{eqnarray}
can become an important
model independent source of information
about the axial and vector strange form factors
\cite{KM-EK,ABGM95}.

The one-nucleon matrix element
of the hadronic neutral current
has the following general form:
\begin{equation}
\left\langle
p'
\left|
J^Z_{\alpha}
\right|
p
\right\rangle
=
\bar{u}(p')
\left[
\gamma_{\alpha}
F_V^Z(Q^2)
+
\dfrac{i}{2M}
\,
\sigma_{\alpha\beta}
q^{\beta}
F_M^Z(Q^2)
+ 
\gamma_{\alpha}
\gamma_5
F_A^Z(Q^2)
\right]
u(p)
\;.
\label{20}
\end{equation}
Here
$ Q^2 \equiv - q^2 $,
where
$ q = p' - p $,
$p$ and $p'$
being the momenta of the initial and final protons,
respectively.
Using the isotopic SU(2)
symmetry of the strong interactions,
for the vector and axial NC form factors
we have
\arraycolsep=0pt
\begin{eqnarray}
&&
F_{V,M}^Z
=
F_{1,2}^3
- 
2 \sin^2\theta_W F_{1,2}^{p(n)}
-
\dfrac{1}{2}
F_{V,M}^s
\;,
\label{21}
\\
&&
F_A^Z
=
\dfrac{1}{2}
F_A
-
\dfrac{1}{2}
F_A^s
\;.
\label{25}
\end{eqnarray}
Here
$ F_1^{p(n)} $
and
$ F_2^{p(n)} $
are the Dirac and Pauli electromagnetic
form factors of the proton,
\begin{equation}
F_{1,2}^3
=
\dfrac{1}{2}
\left(
F_{1,2}^p
-
F_{1,2}^n
\right)
\label{23}
\end{equation}
are the isovector form factors of the nucleon,
$F_A$
is the CC axial form factor,
and
$F_{V,M}^s$
and
$F_A^s$  
are the form factors
that characterize the one-nucleon
matrix elements
of the vector
$ \bar{s} \gamma_{\alpha} s $
and axial
$ \bar{s} \gamma_{\alpha} \gamma_{5} s $
currents.

It is possible to obtain information
about the axial form factor
$F_A$ 
from the investigation of the
quasi-elastic CC processes
\arraycolsep=0pt
\begin{eqnarray}
&&
\nu_\mu + n \to \mu^{-} + p
\;,
\label{12}
\\
&&
\bar\nu_\mu + p \to \mu^{+} + n
\;.
\label{13}
\end{eqnarray}
However,
the axial form factor cannot be determined
from the existing CC data
with an accuracy sufficient to
obtain the value of the constant
$g_A^s$
from the NC elastic scattering data\cite{Garvey}.

In order to avoid
the uncertainties connected with
our lack of knowledge of the axial form factor
$F_A$
we have considered\cite{ABGM95} the
asymmetry\cite{PW73,BK89}
\begin{equation}
\mathcal{A}_{N}(Q^2)
=
\dfrac{
\left(
\dfrac
{ \mathrm{d} \sigma }
{ \mathrm{d} Q^2 }
\right)_{\nu N}^{\mathrm{NC}}
-
\left(
\dfrac
{ \mathrm{d} \sigma }
{ \mathrm{d} Q^2 }
\right)_{\bar\nu N}^{\mathrm{NC}}
}{
\left(
\dfrac
{ \mathrm{d} \sigma }
{ \mathrm{d} Q^2 }
\right)_{\nu n}^{\mathrm{CC}}
-
\left(
\dfrac
{ \mathrm{d} \sigma }
{ \mathrm{d} Q^2 }
\right)_{\bar\nu p}^{\mathrm{CC}}
}
\;,
\label{11}
\end{equation}
where
$
\left(
{ \mathrm{d} \sigma }
/
{ \mathrm{d} Q^2 }
\right)_{\nu N}^{\mathrm{NC}}
$
and
$ \displaystyle
\left(
{ \mathrm{d} \sigma }
/
{ \mathrm{d} Q^2 }
\right)_{\bar\nu N}^{\mathrm{NC}}
$
are the cross sections of the
processes (\ref{091}) and (\ref{092})
and
$ \displaystyle
\left(
{ \mathrm{d} \sigma }
/
{ \mathrm{d} Q^2 }
\right)_{\nu n}^{\mathrm{CC}}
$
and
$ \displaystyle
\left(
{ \mathrm{d} \sigma }
/
{ \mathrm{d} Q^2 }
\right)_{\bar\nu p}^{\mathrm{CC}}
$
are the cross sections of the processes
(\ref{12}) and (\ref{13}),
respectively.
Taking into account only the terms which
depend linearly on the strange form factors,
for the asymmetry we have the following expression:
\begin{equation}
R(Q^2)
\,
\mathcal{A}(Q^2)
=
1
-
\dfrac{ F^s_A(Q^2) }{ F_A(Q^2) }
-
\dfrac{ 1 }{ 8 \left| V_{ud} \right|^2 }
\,
R(Q^2)
\,
\dfrac{  G_M^s(Q^2) }{ G_M^3(Q^2) }
\;,
\label{33}
\end{equation}
where the quantity
\begin{equation}
R(Q^2)
=
\dfrac
{ 4 \left| V_{ud} \right|^2 }
{
1 - 2 \sin^2\theta_W \,
\dfrac{ G_M^{p}(Q^2) }{ G_M^3(Q^2) }
}
\label{34}
\end{equation}
is determined by the ratio of the magnetic form factors
of the neutron and proton
(in Eqs.(\ref{33}))and.(\ref{34})
$\theta_W$ is the Weinberg angle
and
$V_{ud}$
is the element of the CKM mixing matrix).

The electromagnetic form factors
satisfy the approximate scaling relations
\begin{equation}
\begin{array}{l} \displaystyle
G_M^p(Q^2)
=
\mu_p
\,
G_E^p(Q^2)
\;,
\\ \displaystyle
G_M^n(Q^2)
=
\mu_n
\,
G_E^p(Q^2)
\;,
\end{array}
\label{37}
\end{equation}
where
$ \mu_{p(n)} $
is the total magnetic moment
of the proton (neutron) in nuclear Bohr magnetons.
In the scaling approximation we have
\begin{equation}
R(Q^2)
\,
\mathcal{A}(Q^2)
=
1
-
\dfrac{ F^s_A(Q^2) }{ F_A(Q^2) }
-
1.11
\,
\dfrac{ G_M^s(Q^2) }{ G_M^3(Q^2) }
\;
\label{39}
\end{equation}
with a constant
$ R(Q^2) = 8.46 $.
From Eq.(\ref{39}) it can be seen
that the asymmetry has the same sensitivity
to the axial and vector strange form factors.

From Eqs.(\ref{33}) and (\ref{39})
it is clear that
a measurement of the asymmetry
$\mathcal{A}(Q^2)$
could allow to obtain model independent information
about the strange axial and vector
form factors of the nucleon.
Such information would be of great theoretical interest.

New neutrino experiments
(CHORUS\cite{CHORUS},
NOMAD\cite{NOMAD},
ICARUS\cite{ICARUS},
MINOS\cite{MINOS},
COS\-MOS\cite{COSMOS},
etc.)
aimed to search for neutrino oscillations
are taking data or are under preparation.
We think that now
it is a proper time
for considering the possibility
of using these neutrino facilities
in order to get information
on the NC neutrino
(and antineutrino) elastic scattering on protons.

In Ref.\cite{ABGM95}
we have discussed in detail
the uncertainties connected with our knowledge
of the electromagnetic form factors.
Nuclear effects will be considered
in forthcoming publications.

\end{document}